# LEARNING WITH DIFFERENTIAL PRIVACY

Poushali Sengupta[1], Sudipta Paul[2,3], Subhankar Mishra[2,3]

[1] *University of Kalyani, Kalyani, Nadia, West Bengal – 741235*

[2] *National Institute of Science, Education and Research Bhubaneswar*

*Odisha, India – 752050*

[3] *Homi Bhaba National Institute, Anushaktinagar, Mumbai – 400094, India*

[1] tua.poushalisengupta@gmail.com , [2] sudiptapaulvixx@niser.ac.in , [3] smishra@niser.ac.in

## INTRODUCTION

Humans gain "knowledge" by inference from raw events, incidents or structured phenomenon. That implies, as long as it is not meaningful or inferred properly, this "raw data" doesn't become "information" to be inferred that help humans to grab "knowledge" from. Our chapter refers to the definition of knowledge given by (Davenport et al., 1998). It states that from the perspective of an expert with respect to the particular experiences and principles in the right context give a proper structure to asses and integrate new raw data and information. This structure helps to make the transition of the data to "knowledge" - inside a proper intelligent mind. This phenomenon is equally applicable in the daily routine of any organization, processes, norms and practices.

Raw data goes through processes to add contextual meaning in background to become "information". These processes are heavily prone to defect and danger depending on the nature of the data, its sensitivity and its usefulness towards the organizations. One of the biggest dangers that these "information" and "raw data" can face is "leakage". Leakage defined as - when an adversary knowingly disclose sensitive information for business purpose to harm an individual, a community or a particular target for his/her own personal satisfaction. The "leakage" can happen in the pre-processing as well as in the post- processing of "information".

These situations are not desired by any means. Some of the probable solutions are -
- Make sure all endpoints have basic Cyber security systems.
- Use a data backup and recovery solution using encryption-decryption system.
- Clean up the data Storage on the IT Assets after a certain time - window
- Limit user access privileges to only what is absolutely necessary
- Provide Cyber security awareness training to the employees
- *Build security system inherited from the data itself that doesn't need any kind of third-party affiliation, but robust and fast enough to provide enough privacy and security promise simultaneously.*

All of the promises except the last one needs some audit from a 3rd party who or which can be a potential attacker. Also, these promises need extensive monitoring from a human perspective all the time which is a tough and cumbersome work. The last point in the above solutions is formally known as "differential privacy" that is currently the default trend of privacy solution.

The definition of "differential privacy" (DP) will be discussed in section 4 thoroughly. But as a promise, DP might be thought of as a restriction which filters the leakage of sensitive statistics at the time of the publication of aggregated information, with respect to a database, in the algorithmic level. To elaborate the above promise some examples are discussed below (Wikipedia – Differential Privacy),

Government departments and agencies use DP algorithms at the time of publishing demographic and other types of statistical aggregated analysis report with the assurance of the confidentiality of the survey takers and responses,

Companies use DP algorithms at the time of collecting user behavior, in every step to stop leakage. This measure is also applicable to the internal analysts.

In whatever way the sales numbers of a business are covered in the process of hiding, those numbers might appear when the same process will be done in the total calculation of a vast region that the business belongs to with a combinations of addition and subtraction. DP algorithms diminish those possibilities from the root itself even if the attackers use robust, interactive query system.

The research and finding timeline that results in differential privacy is following -

1950s -1960s: Statistical agencies started using electronic information processing system that resulted in the increase of number of data tables as well as potential attack.

- 1977: Statistician Tore Dalenius proposedcell suppression mathematics for statistical disclosure control.
- 2003:Computer scientists Kobbi Nissim and Irit Dinur presented the Fundamental Law of Information Recovery, and its key characteristics.
- 2006: Theoretical computer scientists Cynthia Dwork, McSherry, Kobi Nissim and Smith introduced the concept of DP, with a mathematical definition for the privacy loss and utility associated with any result related to a statistical database (Dwork, 2006; Dwork & Roth, 2014). They proposed (Î, d) DP where the whole processes of statistical functions running on the database are not overly dependent on an individual's data.

The different attack models, reasons to justify them, probable solutions in academia and industry are the main agenda of this chapter which are discussed thoroughly in the following sections.

## DIFFERENTIAL PRIVACY

From the above discussion in the Introduction section it is evident that differential privacy comes with respect to the continuous development in the hope of reduction of information leakage. In the following three subsections, we will discuss about what differential privacy does not promise, different promises of differential privacy, and some useful definitions.

### Problems Regarding Privacy - Preserving Data Analysis

There are some important problems for privacy-preserving data analysis which will be discussed here:

- **Anonymization**
  
  Only anonymization of datasets cannot create the strong privacy. In an anonymized rich database, data enables "naming" an individual by the combination of the birth date, zip code, age etc. to create the uniqueness of the individuals. This "naming" can be used for linkage attack to match an anonymized dataset to a non-anonymized data set. In the year 1997, a linkage attack on Massachusetts hospital discharge dataset with public voter dataset was done by Prof. Latanya Sweeney and name, address, phone number of each persons were leaked.

- **Re-identification**
  
  Re-identification is undesirable and risky because it reveals not only the membership of the database but also the compromising data records of the individuals of the dataset.

- **Large Query**
  
  It cannot be predictable how large query would be. A set of queries or a large query can leak the personal information of individual of a particular database. For example, let M and N are two events and the a set of queries is "How many people are involved with M?"; "How many people are involved with both M and N?"; "How many people are involved with M for last 2 weeks?" and "How many People are involved with both M and N for last 3 weeks?". By asking these questions, one can get the particular answer that he/she wants to know. This situation can compromise the privacy of the individuals.

- **Query Auditing**
  
  In some cases, one can send two or more than two such queries whereby the difference of the answers of these queries, he/she can get the actual answer that he/she wants to know. Privacy can be compromises in this way.

- **Just a few**
  
  In some cases, the data holds privacy in such a way that it does not compromises the private data of all individuals except "just a few" information of individuals. Here, the data base is not fully protected.

- **Ordinary fact**
  Revealing ordinary fact like number of times a person goes to the supermarket in a week, can compromises sensitive information if it is followed over time.

In order to give a plausible solution to the above issues with the existing privacy-preserving data analysis, DP as a concept was plotted by Cynthia Dwork et al. in their 2006 work (Dwork, 2006). "Differential privacy"(Dwork & Roth, 2014) is a methodology by which public sharing of information regarding any dataset is restricted to describe the groups in the dataset but not any information about the individuals. DP addresses a paradox of learning where one can know about the useful information of the given database without accessing the particular sensitive information of individuals. More fundamentally, this provides a facility of learning overall forest data without knowing the individual trees containing the private information.

## The Promise of Differential Privacy

The Fundamental Law of Information Recovery from introduction gives a start point to understand DP where they posed a restriction on the amount of query that might be asked by the analyst to not to reveal any private information by giving an overly accurate answer. DP possess an assurance inherently by the data curator that an individual is not prone to attack, adversely or otherwise, by providing their data for the purpose of use in any kind of research or survey, irrespective of connection to any other databases or data sources available at any corner of the world.

If a medical database is considered for analysis, an insurance company's view of an alcoholic's long-term medical costs can be improved as they can know from the database without accessing individuals' data records that drinking too much alcohol causes various liver disease, most importantly cancer. As a result, one's insurance premium may rise, if the insurer knows that the particular person drinks alcohol. Here the medical database is deferentially private as the information of individuals are not "leaked" to the insurance company and it makes an impact on the participants in the survey for giving answer independently whether he/she drinks alcohol or not.

DP is not an algorithm, it is a theory that ensures, a stream of outputs in arbitrary sequences of choice (responses to queries) is "essentially" equally likely to occur where this term "essentially" is maintained by a privacy parameter. The smaller the value of this parameter the better privacy occurs. This parameter is denoted by $\epsilon$.

## Different Types of Differential Privacy

There are two types of DP. They are discussed below :
- **Local Differential Privacy (LDP):**
  A technique introduced in 1965 (Warner, 1965) named as - "Randomized Response" is the basis of this kind of DP. The users answer the queries using only coin toss probability set. The main plus point of this simple model is that the distribution of the data is always stable enough even when a user suddenly changes the response out of the blue. There is no need to have the affiliation of a third party authorization too. As it always gives the answer for a forest of data instead of a single tree maintaining the above assumption, it is highly adopted to the industry as well as in the academia research.
- **Central Differential Privacy (CDP):**
  This whole technique is depended on the trust and proficiency of the data curator (DC). The DC add random noise to the aggregated answer in the central part after collecting all the data from the local servers. In the local servers the answer to the queries are independent of each other and do not know other servers' identity. Therefore all the answers are individually part of the centralized dataset. As the answers are independent and the DC is adding external random noise, the whole phenomenon leads to CDP.

Therefore, an easy comparison can be inferred from the above discussion which is given in the Table 2.

*Table 2. Comparative discussion on LDP and CDP*

| LDP | CDP |
|---|---|
| No place for a trusted curator, or any data storage. | The data is collected from servers which are independent to each other and stored in a trusted centralized database. |
| Assurance of report production entirely depended on the trusted individual clients. They give their responses using the randomized response technique on the coin toss set. | The curator releases the final report applying the Laplace noise randomly drawn from Laplace distribution for hiding the presence or absence of an individual. |
| Example: RAPPOR in google chromium project for google chrome search engine. | PROCHLO implementation of ESA, ESA revisited, Amplification by shuffling etc. techniques. |

A short timeline regarding the most important inventions and discoveries in DP is given in the following figure 1.

*Figure 1. Short timeline of DP*

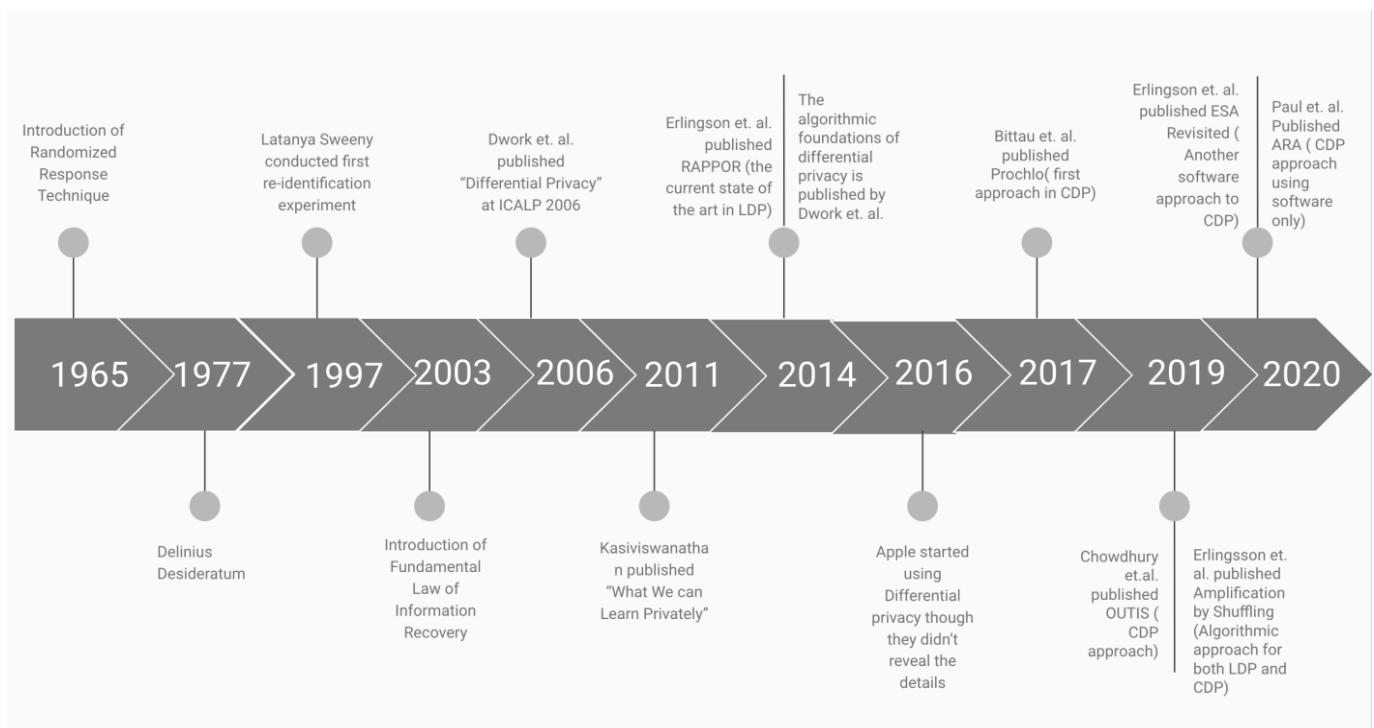

# STATE OF THE ART DP TECHNIQUES ACCORDING TO TIMELINE

Differential privacy is a concept which possesses a huge promise towards security. Here we will discuss about some related works in the industry as well as in the academia using differential privacy such as "what can we learn privately", RAPPOR, PROCHLO, OUTIS, ESA, ARA etc. and compare them with each other.

## Learning with limit

Learning problems form an important section of various computational tasks that are applied by a large number of computational researchers in real life. To recover these types of problems and to find the answer of "what can we learn privately"(Kasiviswanathan et. al, 2011), in this work they explored different approaches of learning concept classes as background theory, that eventually let them propose an algorithm in terms of samples, computations, time and interaction that holds the following promises of improvement.

- Occam's razor's private version.
- Learning privately with an efficient learner.
- Local (randomised response) and SQ learning are equivalent.
- Local learning in two different forms i.e. interactive and non-interactive.

In the following subsection we will go through the preliminary background theories and theoretical result they derived in the process.

### *Preliminary Background Theories*

The preliminary background theories of this work have explored differential privacy, agnostic learning, Probabilistically Approximately Correct (PAC) learning, SQ learning and their efficiency measure. The background of differential privacy will be discussed in the section 4. PAC model is the learning model that has the capability to access a polynomial number of labeled answers. SQ model do not access the answers directly whereas the learner can specify some properties on the example for which an estimate is given up to additive small error, of the probability that a randomly chosen example which follow the distribution D, satisfy the property. The PAC learning is stringently stronger than SQ learning. A brief discussion of PAC, SQ and Agnostic learning is given in the Table 3 below.

*Table 3: Description and mathematical formulation of different learning models.*

| Learning Theories | Description | Formulas and mathematical expressions |
|---|---|---|
| PAC (Probabilistically Approximately Correct) | A concept class depending on some definite distribution will be PAC learnable using a hypothesis class if there exists such algorithm maintaining a polynomial time, whose output will produce a definite hypothesis from the aforesaid hypothesis class maintaining the bounds for probability of failure. | $\forall d \epsilon \mathbb{N}, \forall c \epsilon C_d, \forall \mathcal{X} \epsilon X_d$, an algorithm $\mathcal{A}$ maintaining polynomial in $d, 1/\alpha, \log(1/\beta)$, will give an output from the hypothesis class $h \epsilon \mathcal{H}$, satisfying $$\Pr[error(h) \leq \alpha] \geq 1 - \beta$$ over the drawn examples $z_i = (x_i, c(x_i))$ |
| Agnostic Learning | It is identical to PAC theory with two more addition regarding distribution and bounds for probability of failure. | $\forall d \epsilon \mathbb{N}, \forall c \epsilon C_d, \forall \mathcal{X} \epsilon X_d \times \{0,1\}$, an algorithm $\mathcal{A}$ maintaining polynomial in $d, 1/\alpha, \log(1/\beta)$, , will give an output from the hypothesis class $h \epsilon \mathcal{H}$, satisfying $$\Pr[error(h) \leq OPT + \alpha] \geq 1 - \beta$$ over the drawn examples $z_i = (x_i, c(x_i))$ |
| Private PAC Learning | Same as PAC learning with the addition of privacy parameter $\forall \varepsilon > 0$ on the algorithm $\mathcal{A}$ and polynomial runtime of $\mathcal{A}$. | $\forall d \epsilon \mathbb{N}, \forall c \epsilon C_d, \forall \mathcal{X} \epsilon X_d, \forall \varepsilon > 0$, an algorithm $\mathcal{A}$ maintaining polynomial in $d, 1/\alpha, 1/\varepsilon, \log(1/\beta)$, , will give an output from the hypothesis class $h \epsilon \mathcal{H}$, satisfying $$\Pr[error(h) \leq \alpha] \geq 1 - \beta$$ over the drawn examples $z_i = (x_i, c(x_i))$ |
| Private Agnostic Learning | Same as Agnostic Learning with the addition of privacy parameter $\forall \varepsilon > 0$ on the algorithm $\mathcal{A}$ and polynomial runtime of $\mathcal{A}$. | $\forall d \epsilon \mathbb{N}, \forall c \epsilon C_d, \forall \mathcal{X} \epsilon X_d \times \{0,1\}, \forall \varepsilon > 0$, an algorithm $\mathcal{A}$ maintaining polynomial in $d, 1/\alpha, 1/\varepsilon, \log(1/\beta)$, , will give an output from the hypothesis class $h \epsilon \mathcal{H}$, satisfying $$\Pr[error(h) \leq OPT + \alpha] \geq 1 - \beta$$ over the drawn examples $z_i = (x_i, c(x_i))$ |
| SQ (Statistical query)oracle | It takes a statistical query as an input over some distribution on labeled data with a promise to maintain the output within the tolerance of the expectations of the statistical query drawn from the distribution. | $\left| \mathcal{O}_\mathcal{D}^\tau(\phi) - \mathbb{E}_{(x,y) \sim \mathcal{D}}[\phi(x,y)] \right| \leq \tau$ |
| SQ learning | A class of functions over a definite dimension will be SQ learnable using SQ oracle if there exists such algorithm maintaining a polynomial time, whose output will be produced as an element of the hypothesis class by asking a limited number of queries in the limit of tolerance parameter with a definite probability. | With probability: $1 - \beta$, Output a hypothesis $f \epsilon C$ maintaining: $err(f, \mathcal{D}) \leq min_{f^* \epsilon C} err(f^*, \mathcal{D}) + \alpha$ |
| Local Learning | It is identical to PAC learning with the addition of the SQ learning. | Tolerance parameter $\epsilon$ (0,1) Input function: $SQ_{c,\mathcal{X}}$ Output v maintains the SQ learning promise. |
| Efficient private learner for PARITY | It maintains PAC learn ability. The new introduction here is a class of PARITY functions instead of a class of concepts with a definite failure probability of $\frac{1}{2} + \beta$ | Given, $c_r, r \epsilon \{0,1\}^d$, satisfies: $$\Pr[\mathcal{A}(n, z, \varepsilon) = error(h) \leq \alpha] \geq \frac{1}{2} + \beta$$ |

| MASKED-PARITY | It maintains the weak SQ learn ability with polynomial number of queries, error bounded below $\frac{1}{2}$ | This learning concept class is considered when the distribution $X_d$ is uniform over binary strings with length $d + \log d + 1$ |

In the following table 4 the symbols and description for all the aforementioned learning models is given.

*Table 4. Symbols and Descriptions for the different learning models*

| Symbols | Description |
|---|---|
| C | Universal concept class: $\{C_d\}_{d \epsilon \mathbb{N}}$ |
| $C_d$ | The class of concepts from $X_d$ to $\{0,1\}$ |
| D | $X_d \times \{0,1\}$ |
| $\mathcal{X}$ | $\{X\}_{d \epsilon \mathbb{N}}$ |
| $X_d$ | The distribution of sample $d \epsilon \mathbb{N}$ |
| N | Set of natural number |
| X | Set of all distributions on $X_d$ |
| H | A class of hypothesis. |
| h | A hypothesis that belongs to H |
| error(h) | $P_{x,y \sim \mathcal{D}}[h(x) \neq c(x)]$ |
| c | A specific concept that belongs to $c \epsilon C_d$ |
| β | Bounds of the probability of failure |
| α | Desired error |
| z | Learning algorithm's input. |
| c(x) | A particular element of a particular concept. |
| A | An algorithm |
|  | privacy budget. |
| $c_r$ | A class of parity functions $\{0,1\}^d \to \{0,1\}$ |
| Cr(x) | A set of cross products of random x |
| r | $\{0,1\}^d$ |
| β0 | Probability failure bound for z0 |
| z0 | Neighboring input of z |
| A∗ | An algorithm is found by PAC learning. |
| SQD | A statistical query over the distribution D |
| $SQ_{c,\mathcal{X}}$ | The statistical query oracle that takes as a input function g: $\mathcal{X} \times \{+1,-1\} \to \{+1,-1\}$ |
| τ | Tolerance Parameter and $\tau \in (0,1)$ |
| $\phi$ | Statistical Query function: $\phi : \mathcal{X} \times \{0,1\} \to [0,1]$ |
| $\mathcal{O}_\mathcal{D}^\tau$ | Statistical query oracle |
| n | $O(\frac{\log(1/\beta)}{\epsilon \alpha}(d + \log \frac{1}{\beta}))$ no. of examples |
| $\epsilon$ | Privacy Budget |

For the purpose of better understanding the above discussion the classification of the learning approaches is given in table 5 below. The star (*) marked approaches are the authors own contribution towards different learning approaches.

*Table 5. Associated Works with PAC and SQ learning*

| Learning Approach | Related Learning Theories |
|---|---|
| PAC learning | PAC learning, Agnostic learning, Private PAC learning, Private Agnostic learning, Efficient private learner for PARITY *. |
| SQ learning | Statistical Query (SQ) oracle, SQ learning, Local learning, MASKED-PARITY* |

## RAPPOR (Randomized Aggregatable Privacy-Preserving Ordinal Response)

RAPPOR (Erlingsson et. al., 2014) allows the overall client data to be studied without giving permission for the possibility to access the individual information, with strong privacy and great utility guarantee using randomized response technique in bloom filter. The main contributions of this work are:

- RAPPOR provides a local deferentially private model.
- It generates report by creating noise by applying bloom filter.
- It is the first industrialized implementation of differential privacy.
- it is a fast framework that provides strong privacy with great utility.

The RAPPOR algorithm is in the following flow-chart in figure 2:

*Figure 2. Flow chart of RAPPOR*

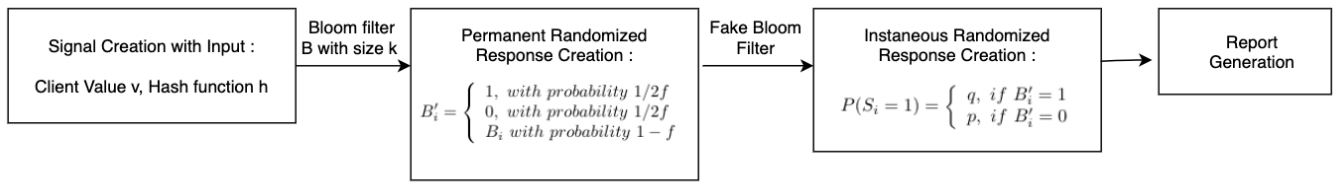

The three types of RAPPOR are explained in the Table 6. In Table 7 all the symbols and description of RAPPOR is defined.

*Table 6. Explanations of three types of RAPPOR*

| One-Time RAPPOR | Basic RAPPOR | Basic One-Time RAPPOR |
|---|---|---|
| As the client itself imposed the one-time collection in this case, the instantaneous randomized response step can be skipped here where the direct randomization on the client's true value is robust enough to give the protection against longitudinal privacy. | It is a special case of RAPPOR. If the report collected are small in size and well-defined enough it can be mapped to a single bit without using a Bloom Filter, reducing the number of hash functions in exchange. An easy example can be data on clients' binary preferences where an "yes" means 1 and a "no" means 2. Here the number of effective bloom filters is 1. | This is the combination of One-time RAPPOR and Basic RAPPOR. A randomization step followed by a settled mapping to unique bits. |

*Table 7. Symbols and Description for RAPPOR*

| Symbols | Description |
|---|---|
| $h$ | Client Value |
| $B$ | First Applied Bloom filter |
| $k$ | Size of the response bit string |
| $B'$ | Permanent Randomized Response |
| $S$ | Instantaneous Randomized Response |
| $f, p, q$ | User tunable parameters |

*Experiments in RAPPOR*

By recall precision graph it is seen that using of only two hash functions (when k and m are fixed) result in better utility as decrease in hash function is the cause for increase in expected recall.

For the experiment, two simulated and two real world data were used to apply RAPPOR algorithm. The first one of simulated data was used for Basic One-time RAPPOR and the underlying distribution was found to be normal distribution with enough noise. The second simulated example was used to apply modified RAPPOR for collecting strings and it showed the exponential distribution of string frequencies. The third example was real world data on processes running on windows machines. In this case, the frequency of a particular process "BADAPPLE.COM" was estimated with 128 bit size bloom filter, 2 hash functions, 8 cohorts where q = 0.75, p = 0.5 and f = 0.5. The estimated frequency of BADAPPLE.COM was found to be 2.6%. The data on Chrome Homepages were used for last example and it was observed that the analysis of RAPPOR can discover URL domains of homepage, with statistical confidence, if their frequencies exceed 0.1% of the responding population.

## PROCHLO

It's a common practice now a day to monitor the user's software activities thoroughly to provide good and elevated service by the companies. PROCHLO (Bittau et. al., 2017) describes architecture of a principled system - Encode, Shuffle, analyze (ESA) to perform such activities by maintaining higher *utility* through preserving strong privacy guarantee. Here a pipeline is used where a randomly drawn sample from users' input is (Encoding :) encrypted first by pushing noise randomly, encoded by breaking the whole encrypted string into data fragments, then (Shuffling:) shuffled by a secret shuffler and at last (Analyzing:) the report is generated from the sample.

*Architecture: PROCHLO*

It uses the Google SGX as the secret shuffler. SGX has small memory that cannot hold the millions of data, and that's why rather transferring the whole sample of users' input directly to the shuffler, PROCHLO creates crowd IDs. Let, the total number of users' input is N and the number of crowd ID is B+. Then the number of input having the same crowd ID is $D^+ = \frac{N}{B^+}$. After getting the bunch of inputs belonging to same crowd ID, they are divided into fragments for encoding.

- **Step 1: Encoding**

  The encoding can be done by t∗ secret shares where t∗ secret shares split a secret S∗ in a filed F into arbitrary S∗1, S∗2,.... so that any t∗ −1 shares can hide all information about S∗. A secret share of S∗ is tuple [x,ρ(x)] for randomly chosen nonzero x ∈ F where S∗= ρ(x). Let t secret sharing encode of an arbitrary string m with parameter t∗ is a pair as (c,aux) where c is the cipher text which is a deterministic encryption under a key $K_m$= H(m) and aux be the t∗ secret shares of $K_m$. But the m is recoverable by using robust decryption attack with the help of aux₁, aux₂, ....auxₜ to find the $K_m$ and by using $K_m$ he/she can decrypt c and in this way one can create an attack to disclose the private data of individuals.

So, PROCHLO use the data fragmentation for encryption. For example, consider an analysis on a database based upon the audiences' ratings on movies. In this database each entry is supposed to be "individual" and a concern of privacy. To protect privacy, fragmentation is used here. consider a movie rating set is $\{(m_0, r_0), ((m_1, r_1), (m_2, r_2))\}$. Where $m_i$ (i=0:2) is the name of the ith movie and $r_i$ (i=0:2) is the rating of the ith movie. Now, the elements of this rating set can be encoded as their pairwise combinations like $\langle(m_0, r_0), (m_1, r_1)\rangle$, $\langle(m_0, r_0), (m_2, r_2)\rangle$ and $\langle(m_1, r_1), (m_2, r_2)\rangle$ where each pair will be transmitted to the secret shuffler for the independent shuffling. The shuffler is trusted but always curious about the private data of individuals. Here, for generating the crowd IDs and encryption, it is hard for the shuffler to guess about the private information of any particular individual.

- **Step 2: Shuffling**
In the next part of this pipeline, this shuffler performs four tasks and those are anonymization, shuffling, thresholding and batching.

  o Anonymization - Anonymization is created by striping metadata which is not enough for privacy.

  o Shuffling- For this, the SGX do the oblivious shuffling. Based on Melbourne shuffle technique, PROCHLO introduces the Stash shuffle that actually improves the all previous problems. Here, consider B∗ input and outputs buckets each having $D'' = \frac{N}{B*}$ items. At first, $D''$ items of each input buckets are random shuffled with $B*-1$ bucket separator. The shuffles determine which items will fall into which target output bucket. Then for every output bucket, as long as there is still room in the maximum C items to output, the items from input bucket are read, decrypted and deposited to the output bucket. If the bucket is full of C items, then the next items will go to the Stash and start waiting for the next target bucket. Finally, if some output bucket still not filled with C quota, then they are filled with dummy variables. After all input buckets has been processed, the Stash may leave over with some items. These items are drained by filling extra K items per output buckets. At the end of this phase, K is set to be $\frac{S*}{B*}$ where S∗= stash space and B∗= bucket number. Now, all the items in output buckets are re-encrypted, shuffled and dummy items are filtered out from every bucket. Each bucket drops $d''$ item in this phase. After that, $D''$ items are forwarded as output at a time.

    ▪ The phase proceeds in a sliding window w buckets of intermediate items. At the time of shuffling, there are two secret shufflers. The encoder computes $(g^r, h_*^r, \mu)$ where (g,h∗) is the key of shuffler 2 and r is a random value. Shuffler 1 is a blind shuffler that creates a secret $\alpha'$ for each tuple where $\alpha' \epsilon Z_p$ and compute $(g^{r\alpha'}, (h_*^r \cdot \mu)^{\alpha'})$
    ▪ After that, the shufflers 1 batch the data, shuffles and forward the blinded items to shuffler 2. Shuffler 2 uses its private key that is the secret x∗ such that h∗= gx∗ on input (u,v) to compute $v^{\frac{u}{x*}}$ and recover μα0 = $H(crowdID)^{\alpha'}$ It works with crowd ID that are already hashed and has raised to a secret power $\alpha'$. With the blind crowd ID, shuffler 1 cannot do the dictionary attack, since it does not know the secret key of the shuffler 2, whereas shuffler 2 also cannot do such attack as it has not any idea about the secret $\alpha'$ of shuffler 1. So, at the time of shuffling, privacy is maintained.

  o Thresholding- Even stripped and shuffled data may identify a client by uniqueness and to prevent this the shuffler do thresholding by generating crowdID.

  o Batching- After that, the shuffler forwards the data to the analyser infrequently, in batches.

- **Step 3: Analyzing**
The analyzer decrypts, stores and aggregate those data comes from the shuffler. The analyzer's output is considered public. Analyzer uses the PINQ, FLEX or the systems such as Airavat to maintain the final protection of user's privacy.

All the symbols and description of PROCHLO are given in the following table 8:

*Table 8: Symbols and Description of PROCHLO*

| Symbols | Description |
|---|---|
| $S_*$ | Secret string. |
| $t^*$ | Number of secret shares. |
| F | Field contains all the secret strings. |
| $K_m$ | Key. |
| c | Cipher text. |
| $aux_i$ | $i^{th}$ Number secret sharing among all t shares. |
| N | Total number of user inputs. |
| B* | Number of buckets |
| $D''$ | Number of outputs each output bucket has. |
| C | Maximum quota of each target bucket. |
| $S^*$ | Stash. |
| K | Number of extra elements per output bucket. |
| $d'$ | Number of items dropped by each output bucket. |
| (u, v) | Input from users. |
| $B^+$ | Number of crowdID. |
| $D^+$ | Number of inputs having same crowdID. |

*Experiments: PROCHLO*

It performs four experiments with Crowd, Secret Crowd, No-Crowd and Blinded-Crowd to privately learn word frequencies on sample of size 10K,100K and 1M. It is seen that, if data with No-Crowd are transferred to the pipeline, it will have no privacy and highest utility whereas if we pass the data with Crowd , then the privacy will be improved but not so much. By creating secret Crowd, this problem can be recover, but still the result is not satisfactory so much. all of this problem can be removed by applying blinded Crowd to the data which provides strong privacy guarantee with great utility.

PROCHLO provides much better utility than RAPPOR while maintaining strong privacy guarantee. Here, data have less noise than RAPPOR reports, so the utility becomes higher. PROCHLO not only offers good balance between utility and privacy, it introduces both new cryptographic primitives and a new algorithm of oblivious shuffling. PROCHLO is relatively simple, easy to understand system and also has the straightforward realization of the ESA structure that minimizes trust issues.

## Amplification by Shuffling

A important task in data analysis is the monitoring of the statistical properties of the data in a manner that requires repeated computation the entire data set which is involved and that type of monitoring can directly or indirectly harm the data by exposing private information about sensitive attribute of user. In case of centralized differential model, the response should be stable while the neighboring data set differs from n rows. So, the centralized differential model provides better privacy than the LDP model.

Inspired by differential privacy under continuous observation, Amplification by Shuffling (Erlingsson et. al., 2019) provides an algorithm that gives high accuracy online monitoring on users' input in LDP model whose privacy cost is poly-logarithmic in number of changes in users' input and it shows how LDP guarantees about privacy protection while doing online monitoring even when the users report repeatedly, over multiple timesteps, and whether they report on the highly correlated value, on the same value or independently drawn values. This privacy amplification technique shows that any permutation-invariant algorithm satisfying differential privacy will satisfy $O(\epsilon \sqrt{\log\left(\frac{1}{\delta}\right), n})$ central differential privacy.

## Mathematical Model and Lower bound

Let, we want to know the number of times a software is used by n users at the time horizon d and the input is taken repeated time at the time point t [d]. Consider a population of n users reporting a boolean value about their state at each time period $t \epsilon [d]$ where [d]= (d1, d2,.....) and d is the power of 2. $S_{ti} = \{S_{ti}[1], S_{ti}[2], ......, S_{ti}[d]\}$ denotes the ith user's state at the time point $t \epsilon [d]$ and they can change their input at most k times. Let, $x_i = \{x_i[1], x_i[2], ....., x_i[d]\}$ denotes the changes by the ith user at the time point $t \epsilon [d]$. Then, $S_{ti}[d] = \sum_{l \epsilon [t]} x_i[l]$ and $f_t = \sum_{i=1}^{n} x_i[t]$ will be the running count or marginal sum. Now, h is the hth level of the balanced binary tree and Hi be the number of nodes of the hth level of the ith user's tree and it is denoted by $H(h_i) for h \epsilon [\log_2 d+1]$. after that, k and c are initialised as k=0 and c=0, d and k∗ is set up and the input xt is taken. Update $x_t, t, \epsilon$ at each time step and modify k and c where the process is differential private and $t \leq d, x_t \in \{-1,0,1\}$. If $x_t \neq 0$, k= k+1, and when k reaches k∗, c takes the value of xt. Otherwise, if c=0 and t is divisible by 2h-1, u take random value from {-1,1}, else b takes value from $2 \times [Ber\left(\frac{e^{\frac{\epsilon}{2}}}{1+e^{\frac{\epsilon}{2}}}\right) - 1]$ and u takes (b × c). Now for response, the algorithm creates a $T_{sum}\left[h, \frac{t}{2^{h-1}}\right] = \sum_{i,h,t=h} u_{i,t}$ and $\bar{f}_t = \left(\frac{e^{\frac{\epsilon}{2}}+1}{e^{\frac{\epsilon}{2}}-1}\right).k\log_2 d.\sum_{[h,i]} T_{sum}[h,i]$. Here $T[h,i]$ is the sum of independent random variables that come from the range [-1, 1]. Now,

$$\forall [h,i]: |T[h'i] - E[T[h,i]]| \leq c_\epsilon \sqrt{n \frac{\log \frac{2d}{\beta}}{\log_2 d}}, \text{ where, } c_\epsilon = \left(\frac{e^{\frac{\epsilon}{2}}+1}{e^{\frac{\epsilon}{2}}-1}\right)$$

By scaling with $c'_\epsilon k \log_2 d$ and multiplying by log2 d we get,

$$\forall t \epsilon [d]: |f_t - \bar{f}_t| \leq c_\epsilon k (\log_2 d)^{\frac{3}{2}} \sqrt{n \log_2 \frac{2d}{\beta}}$$

Now, the algorithm $A_{ldp}^{(i)}$ and $D = x_{i:n}$ are used where $A_{ldp}^{(i)}: S^{(1)} \times S^{(2)} \times ........ \times D \rightarrow S^{(i)}$ be the input from ith user. After taking input $x_1, x_2,,......, x_n$, a random number I is chosen from [1:n] and then the first element of the data is swapped with the Ith element. Now, the local randomiser is operated with the data set. Let π be any permutation of the set [1:n], then $\pi(D) \leftarrow \{x_{\pi(1)}, x_{\pi(2)}, ....., x_{\pi(n)}\}$. After Local Randomisation, again a shuffling is done where zi takes the value (z1:(i-1), xi) and the generated report is $\{z_{\pi(1)}, z_{\pi(2)}, ....., z_{\pi(n)}\}$.

Let, T is sampled from the distribution of I conditioned on $Z_{1:n} = S_{1:n}$. Then,

$$\frac{P(Z_{1:n}= S_{1:n}|T=i)}{P(Z_{1:n}= S_{1:n}|T=j)} \leq e^{2.\epsilon}$$

Now, we can write

$$P(T = i|Z_{1:n} = S_{1:n}) = \frac{P(Z_{1:n} = S_{1:n}|T = i).p(T = i)}{P(Z_{1:n} = S_{1:n})|} \leq \frac{1}{n}e^{2.\epsilon}$$

Because T is uniform over [n] and $p = p' = \frac{1}{n}$. Applying privacy amplification by shuffling and advanced composition theorem we get

$$n\epsilon(e^\epsilon - 1) \leq \frac{65}{64}n(\epsilon_1)^2 \leq \frac{2}{3}\epsilon_0 \sqrt{\frac{\log \frac{1}{\delta}}{n}}$$

Where, $\epsilon < \epsilon_1 \sqrt{2n \log \frac{1}{\delta}} + n\epsilon_1(e^\epsilon - 1)$ and $\epsilon_1 \leq \frac{8\epsilon}{n}$, $\epsilon_0 \leq \frac{1}{2}$, δ is the small distance between two datasets.

Each set of $n'$ users for which the result is applicable are still be guaranteed by a factor $\sqrt{n'}$ reduction in the centralised privacy model for the worst possible case. Here shuffling after the local randomisation technique satisfies $(\epsilon, \delta)$ differential privacy at index i in the central model where $\epsilon = 12\epsilon_0 \sqrt{\frac{\log \frac{1}{\delta}}{|S|}}$; $S \subseteq |n|$ such that $i, j \epsilon S, A_{ldp}^{(i)} = A_{ldp}^{(j)}$

*Table 9. Symbols and Description for Amplification of Shuffling*

| Symbols | Description |
|---|---|
| t | Time Point |
| d | Time Horizon |
| $S_{ti}$ | $i^{th}$ user's state at the time point t |
| $x_i$ | Changes by $i^{th}$ user. |
| $f_t$ | Running count |
| $\bar{f}_t$ | Expected running count |
| $H(h_i)$ | Number of nodes at the $h^{th}$ level of tree T |
| $T(h,i)$ | sum of independent random variables that come from the range [-1, 1] |
| D | Database |
| $\pi$ | Any permutation of the set [1: n] |
| $\epsilon$ | Privacy budget o the data set. |
| $\epsilon_1$ | Privacy budget of neighboring data set D' |
| $\epsilon_0$ | Privacy budget of $A_{ldp}^{(i)}$ |
| $S^{(i)}$ | Range space of $A_{ldp}^{(i)}$ |
| $A_{ldp}^{(i)}$ | ith local differentially private algorithm |

The result of Amplification by Shuffling is encouraging, and it secures the differential privacy in central differential model and make the stable response for differing single input in two data sets. But this algorithm gives very poor response for longitudinal data. This formalisation assumes that the user population to be static which does not compared to real world.

## OUTIS

This paper provides a framework that provides an accuracy guarantee just like CDP model (Chowdhury et. al., 2019) without any trusted data collector. Here crypt employs two non- colluding semi-trusted server AS (Analytic Server) and CSP (Cryptographic Service Provider )that run differential privacy on the encrypted data given by the data owner.

### Cryptographic Primitives: OUTIS

In their paper they used four cryptography schemes. They are explained in the following –

- Linear Homomorphic Encryption (LHE) - This scheme consists of three stages i.e. Key generation, Encryption and Decryption on the basis of the group $(\mathcal{M}, +)$. In the first stage a user tunable security parameter K is being passed as input in the pipeline which outputs a pair of keys. Those are secret and public keys - $(S_k, P_k)$. In the second stage, using the public key $P_k$ a randomized algorithm encrypts a message $m \in \mathcal{M}$ to $c$. The third and the last part decrypt $c$ from the last stage using $S_k$ to recover the plaintext message $m$ deterministically.
- Labelled Homomorphic Encryption (LabHE) – By the introduction of pseudo-random function every LHE scheme can be changed to LabHE. The LabHE scheme also has the ability to multiply two LabHE ciphers.
- They have used two more primitives too. Those are "$operator \oplus$ "and "Secure computation using garbled circuit". $operator \oplus$ is mainly used for the repetition purpose. In the later scheme, for two private inputs from two parties, no party can learn more than $f(input1, input2)$ for a function $f$. This is an inherent generator scheme using garbled input in every steps.

### Architecture: OUTIS

The algorithm is given below-

- **Setup Phase:**
  The owner initializes the privacy budget $\epsilon^\beta$ for CSP and stores it in CSP's privacy engine module. Then the CSP key manager produces a pair of keys$(S_k, P_k)$. For labHE and publishes $P_k$ stores$S_k$.

- **Data Collection Phase:**
  In this phase, data owner encrypts and encodes his data with the help of data encryption technique and data encoder and then sends the encrypted and encoded data to As and after that, data owner gets offline completely. As, aggregates the data into a singleencrypted database D by aggregator module.
- **Program Execution Phase:**
  Here AS executes data analyst provided crypt programs that accesses sensitive data by various transformation operators like Cross Product, Project, Filter, Count, Group By Count, Group By Count Encoded, Count Distinct etc and measurement operators like Laplace, Noisy Max etc which are DP operations to create noisy answer. Measurement operators require interaction with CSP as they need to decrypt the data and also to check whether the privacy budget is exceeded or not. These functions can be done by the CSP's data decryption and privacy budget modules.

*Table 10. Symbols and Descriptions for subsection OUTIS*

| Symbol | Description |
| --- | --- |
| $\epsilon^\beta$ | Privacy budget for cryptographic service provider |
| $S_k$ | Secret key |
| $P_k$ | Private key |
| c | Cipher text |
| m | Message to be encrypted |
| $\overline{D}$ | Aggregated data |

The first two phases occur exactly once at the beginning and every subsequent program are handled by the corresponding program execution phase.

## *Experiments: OUTIS*

For experiments they use a schema database <Age, Gender, Native Country, Race> and shows 7 cryptographic examples over it. They uses the crypt program p1 to find the cumulative distribution of age where the first step is to compute 100 ranges queries and the i[th] query computes the number of persons having the age $\epsilon$ [0,i] in $\overline{D}$ with privacy guarantee $\epsilon_i$ then they applied a sequence of transformation operators in each range of queries and all of these operators have stability bound 1. For this, the resultant range query has a sensitivity upper bound 1. In this way, the subsequent measurement operator Laplace takes the privacy budget $\epsilon_i$ and sensitivity $\Delta = 1$. After looping over 100 ranges, p1 gives a noisy plain text $\hat{V} = \{\hat{c_1}, \hat{c_2}, ...., \hat{c_{100}}\}$ and at the end of the program the privacy budget is $\sum_{i=1}^{100} \epsilon_i$.

## **ARA (Aggregated RAPPOR and Analysis for Central Differential Privacy)**

The main aim of ARA (Paul et. al., 2020) is trying to produce a bridge to keep best of both the worlds in terms of LDP and CDP, which is a software approach, comparatively less expensive, fast with less complexity in the analysis and overall method.
The main contributions of ARA are following –
- The promises of DP are maintained.
- Fast enough analysis phase (Around 1.5 hours for 1000 users times 100 loops of analysis)
- Correct identification of the highest occurred true value every time with varying achievement.
- Simple probabilistic analysis method in comparison to OUTIS or PROCHLO.

## *Methodology and Experiments: ARA*

ARA collects data after cloning the RAPPOR code from the Google repository of GitHub by running it 100 times to make their own database using the generated RAPPOR reports. It only uses ten true values for convenience where RAPPOR has used 100 true values.

In the first step the sampling is done in 100, 1000, 10000, 20000, 25000 samples taking at a time randomly from the created dataset without any repetition. Then the TF-IDF value is calculated with the formula given in the table 11. The two achievements from this step are –
- No more than $[k/2] + 1$ positions can grab "on bit"s. Here, $k = 32$.
- Constant values for 17 bit positions are calculated for the "on" bits.

These values are now predefined for the next step and securely preserved by the trusted data curator.

*Table 11. The 17 constants value retrieved using continuous sampling on the RAPPOR reports*

| Number of 'on bit' in the string | Constant Value $C_v$ (v ranged from 1 - 17) |
|---|---|
| 1 | 1.20201279 |
| 2 | 1.0927389 |
| 3 | 0.993399 |
| 4 | 0.90309 |
| 5 | 0.80618 |
| 6 | 0.727 |
| 7 | 0.660052 |
| 8 | 0.60206 |
| 9 | 0.550907 |
| 10 | 0.50515 |
| 11 | 0.4637573 |
| 12 | 0.425969 |
| 13 | 0.3912066 |
| 14 | 0.3590219 |
| 15 | 0.329059 |
| 16 | 0.30103 |
| 17 | 0.274701 |

*Source: Paul et. al., 2020*

In the second step, the weighted sum for the prr and irr strings is being calculated using the formulas given in the table. These sums are stored for a minimum time for the purpose of experiment .

The last step is the testing phase, where the RAPPOR reports are being drawn again randomly from the testing portion of the created database and the weighted sum is being calculated and matched against the preserved results in the second step. The formulas of each steps can be consulted from the original ARA (Paul et. al., 2020) paper and the description of the terms used in those formulas are given in the following table 12.

*Table 12. Symbols and Descriptions for ARA*

| Symbol | Description |
|---|---|
| $TF(t", d^*)$ | Term Frequency |
| $t"$ | Term |
| $d^*$ | Document in which the term t" has occurred |
| $f_{t",d^*}$ | Frequency of occurrence t" in d* |
| $IDF(t", d^*)$ | Inverse document frequency |
| $1 + \|d^* \epsilon D^* : t" \epsilon d^*\|$ | Total number of documents in the corpus |

| N | Number of documents where t" has appeared |
|---|---|
| W | Weighted sum |
| $C_{prr}$ | The count of 'on bit' in the prr string |
| $C_{irr}$ | The count of 'on bit' in the irr string |
| V | Cohort value |
| $C_{C_{prr}}$ | Value taken from the $C_v$ in the Table 11 |
| $C_{C_{irr}}$ | Value taken from the $C_v$ in the Table 11 |

The main drawbacks of ARA are that the accuracy is not more than 52.28% on an average and the detection ability of the second major true value is poor.

**Encode, Shuffle, Analyze, Privacy Revisited: Formalization and Empirical Evaluation**
In this paper (Erlingsson et. al., 2020) they tried to bridge all their previous works starting from RAPPOR, PROCHLO – an implementation of ESA (encode, shuffle, analysis) till Amplification by shuffling, with the aim to give a proper algorithm that works equally good in both the LDP and CDP environment with an extensive experimental results using MNIST and CIFER-10 datasets. The key techniques they talked about here are the following -

- **Anonymity**
- **Reporting with the preservation of privacy**
- **Sketch based encryption and its limitation**
- **Cons of fragmentation**
- **Cons of shuffling**

*Background Theorem*
In the proposed algorithm here, data are collected from the users first and then they are encoded by one of the techniques such as one-hot encoding, attribute fragmentation, report fragmentation and sketch-based encoding and all of these schemes produce reports with localized differential privacy guarantee. Each and every encoding system have their own pros and cons too. After completing encoding, the LDP reports then are being shuffled by k-shuffler to aggregate in the next step. After that the algorithm of amplification by shuffling is applied here and the final reports are generated which hold centralized and local differential privacy guarantee equally. The parts of this algorithm is discussed below -

- **One-Hot Encoding**
  The type of encoding has strong impact on utility of the differentially private algorithm. Let D be dictionary of elements and id D is not too large, then a data input x can be encoded by one-hot encoding where each data record x will hold an element in D.But when the D is large enough, it is not a suitable idea to apply one-hot encoding whereas sketching algorithm can be used.

- **Sketch-Based Reports**
  The main idea of this scheme is to reduce a given domain $\{0,1\}^{K'}$ to $\{0,1\}^{k'}$, $k' < K'$ via hashing and then use locally private protocols to operate over the domain of size $k'$. To avoid significant loss due to hashing, it is performed by multiple independent hash function. But it is observed that sketching is not a requirement for practical deployment in regimes with local differential privacy.

- **Attributes Fragments**
  In this case, each data record x is encoded as a binary vector with k or fewer bit set and each k vector coordinate is called attributes where x = Σx$_i$, x$_i$ is one-hot encoded vector. Now, either each xi splitting privacy budget accordingly go through the randomizer R∗ or one of the xi is drawn as the sample and spend the entire privacy budget to send is to R∗ which is a randomly chosen randomizer. There are two types of local randomizer, one holds the algorithm for replacement privacy guarantee and another holds the algorithm for removal privacy guarantee.

- o Replacement LDP
  An algorithm R*:D → S is a replacement $(\epsilon, \delta)$ deferentially private local randomiser if for all S ⊆ S and for all $x, x' \epsilon D: P[R^*(x)\epsilon S] \leq e^\epsilon P[R^*(x')\epsilon S] + \delta$

- o Generalized Removal $(\epsilon, \delta)$ DP
  A randomized algorithm M:$D^n$ → S is a replacement satisfies removal $(\epsilon, \delta)$- differential privacy if there exist an algorithm $M': D^n \times 2^{[n]}$ → S with the following properties:
    - for all $D \epsilon D^n, M'(D, [n])$ is identical to M(D)
    - for all $D \epsilon D^n$ and $I \epsilon [n], M'(D, I)$ depends only on elements D with indices I.
    - for all S⊆ S, $D \epsilon D^n$ and I, I'⊆ [n] where we have that |IΔI'|=1,
      $P[M'(D,I) \epsilon S] \leq e^\epsilon P[M'(D',I') \epsilon S] + \delta$

- o Removal LDP
  An algorithm R*:D → S is a removal $(\epsilon, \delta)$ differentially private local randomiser if exist a random variable R0 such that for all S ⊆ S and for all $x \epsilon D$, $e^{-\epsilon} P[M'(D,I) \epsilon S] \leq e^\epsilon P[M'(D',I') \epsilon S] + \delta$

After sampling each z from $\{x_i : i \epsilon [k]\}$, it is seen that sending each attribute of z independently to LDP randomiser that produce anonymous reports, is advantageous.

- Report Fragments
  Here, a sequence of LDP reports which is generated by multiple independent applications of the randomizer R∗ to x has come and each such report is called report fragment containing less information than the entire LDP report sequence. Thus, it improves privacy guarantee more.

After completing the encoding parts, the fragments are sent to the shuffler. There are k secret shufflers. Each fragment chooses one shuffler randomly and each attribute go to the separate channel of the selected shuffler. Then the secret shuffling is done and shuffle data came where one cannot guess which data fragments come from which shuffler. In this way, the data remains deferentially private. After shuffling the aggregate is calculated and at last amplification by shuffling technique is applied to make the whole process centrally deferentially private.

## USEFUL BACKGROUND ON DIFFERENTIAL PRIVACY
To understand the fundamental theory of DP one need to goes through proper steps. These steps are following:

- **Understanding the background probability theorems to define DP**
  This step consists of learning about probability simplex, randomized algorithm, distance between adjacent databases, definition of differential privacy derived from the afore mentioned concepts and finally the intuition of post processing through the definitions and proposition mentioned. All of these are explained in the Appendix 1.

- **Understanding the tools to build the DP algorithm**
  This step provides the definition, remarks, theorems and claims about probabilistic tools like additive Chernoff bound, multiplicative Chernoff bound, Azuma's inequality, Stirling's approximation. It also helps to understand how to introduce noise at the time of data collection or structured surveys by enlightening us on the concepts of Randomized response, Laplace mechanism, Exponential mechanism etc. All of these are explained in the Appendix 2.

- **Understanding the way to use those tools to build a DP algorithm**
  The final step helps us to actually build a DP algorithm for real purpose by letting us understand about the Composition theorem, its technicalities, definition, advancement, Sparse vector technique, its algorithm, theorems, definition etc. All of these are explained in the Appendix 3.

To understand if a proposed DP algorithm is good or bad, we need to calculate its accuracy and utility (Alvim et. al., 2011). The flow of utility and accuracy inside the DP system is given in the following figure 3. The definitions are explained in Appendix 3.

*Figure 3. Leakage and utility workflow using differential privacy*

*Source: Alvim et. al., 2011*

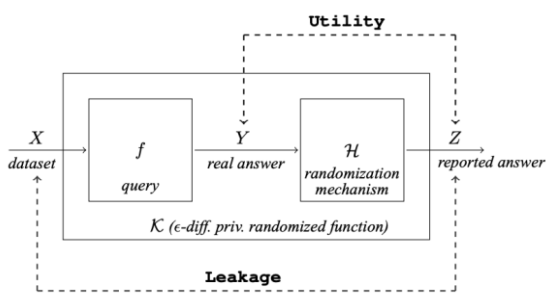

## CONCLUSION AND FUTURE SCOPE

Industries are becoming heavily dependent and invested in data. In a lot of situations, user experience can be improved based on the meta-data we learn from what other users are doing. Few examples are trending words, relevant suggestions, battery hungry websites, popular emojis. However, answering these questions needs personal data of the users.

Differential privacy technique helps us comprehend what users are doing, while protecting the privacy of individual users. It helps us learn about the user groups without learning about the members of the group. It makes sure that data cannot be reproduced even by the organization that is collecting the data. It achieves so with the idea that a biased statistical noise is enough to disguise a user's data and when aggregated over large users, can still give us insight on the community. With increase in cyber-attacks as well as data harvesting; differential privacy is the only ethical way of data collection.

In this chapter, we discuss differential privacy with its history and origin as well as the current state of the art methods. We discuss the architecture or the model of the various algorithms adopting the differential privacy technique along with the required theorems, proofs and experiments to justify their privacy and utility. We believe this chapter will communicate the audience about the current status of this research area as well as engage the readers with exploring research directions in the area of differential privacy.

## ACKNOWLEDGEMENT

This research was supported by Dept. of Science and Technology (Govt. of India).
 Grant no:NRDMS/UG/S.Mishra/Odisha/E-01/2018

## ADDITIONAL READING:

This Section contains some of the current developments on Differential Privacy. They are following:

- **Continuous Release of Data Stream Under Both Centralized and Local Differential Privacy:** In their work (Wang et al.,2019) propose an exponential mechanism with quality function, along with two algorithms ToPS (threshold optimizer, perturber and smoother) and ToPL ( ToPS applied in LDP) which provides the solution regarding the publishing of a stream of real valued data satisfying DP. ToPL is the first LDP algorithm for streaming data and it is the hierarchical version of traditional PAK algorithm (Dwork et al., 2010). ToPL is a method designed with the framework of ToPS for outputting streaming data in LDP. Here, as user perturbs their values before sending to the server, there is no need to trust the server. Threshold optimizer uses frequency estimation instead of applying exponential mechanism (as user has local view) which is less accurate than EM (exponential mechanism). After achieving optimal threshold parameter, the user report is first truncated then the consistent noise is applied. As the reports are unbiased themselves, to answer a range of queries, this algorithm doesn't use the smoother here. But they use the PAK setting in threshold optimizer assuming that the distribution stays same with further information like data changes slowly or regularly, which is a drawback.
- **A Privacy Preserving Randomized Gossip Algorithm via Controlled Noise Insertion:**
This work (Hanzely et al., 2019) introduces a duel algorithm based on average consensus (AC) that provides iteration complexity bound and performs extensive numerical experiments while protecting the information regarding initial value stored in the node. For a particular graph structure, at first the initial value is provided after adding noise to the system for propagating across the network, but in the following iteration , the previously added noise is removed and the new noise with smaller magnitude is applied to ensure the convergence to the true average that maximise the utility of the system obviously.
- **A General Approach to Adding Differential Privacy to Iterative Training Procedures:**
This work (McMahan et al., 2018) introduces an algorithm with the modular approach to minimize the changes on training algorithm by various configuration strategies of privacy mechanism which is applied to heterogeneous set of vectors (like gradients of different layers of DNN, matrices, batch normalization parameters with the different properties). The algorithm provides an integrated privacy mechanism that differs in granular privacy guarantee offered and the method of privacy accounting, where, after choosing the subset from the training data set with a probability parameter, using statistical queries - the privacy mechanism containing noises is applied to each collected aggregates and the accounting procedure is used to compute a final (ε, δ)- differential privacy guarantee with a good balance between privacy and utility. The major drawbacks are, no utility bound is given to justify the good balance between privacy and utility, the distribution is defined over a finite domain, the source of randomness is assumed to be guaranteed and computationally secured and lastly they leave open the task of developing a provable floating point implementation using Laplace mechanism on DP-SGD and iterating it into a ML Library.
- **Capacity Bounded Differential Privacy:**
This paper (Chaudhuri et al.,2019) introduces the capacity bounded differential privacy where the adversary that distinguishes the output distribution is assumed to be not bounded in computational power. This work models adversaries using restricted f divergence between the probability distribution and study various properties of the definition and algorithm that satisfy them. They also show that the Laplace and Gaussian mechanisms are also giving better privacy by applying this new algorithm.
- **BUDS: Balancing Utility and Differential Privacy:**
This work (Poushali Sengupta et al.,2020) introduces a mechanism along with Iterative Shuffling that ensures a good balance between utility and privacy guarantee of the dataset. The whole mechanism can be decomposed into two parts: Application of Query analysis and Iterative Shuffling (IS). The main drawback of this work is: one hot encoding is used for encoding which is a problem for big dimensional data set. Also proper architecture of Query Function is not provided here.

**Key terms and Definitions:**

This section is divided into four parts. The first part contains the key terms and the following three parts contains the necessary definitions and theorems for the learning of differential privacy.

*Key terms:*
PAC Learning, SQ Learning, Private PAC Learning, Private SQ Learning, Randomized Response, Differential Privacy, RAPPOR, ESA, ARA, BUDS

*Definitions and Theorems for the first step:*

In the second part we will discuss some useful definitions and background theorems that work for the first learning step of DP discussed in the useful background on differential privacy section in alphabetical order as instructed by the publisher (Dwork et. al., 2014).

*Differential Privacy*
A randomized algorithm $\mathcal{M}$ with domain $\mathbb{N}^{|\mathcal{X}|}$ is $(\epsilon, \delta)$ - deferentially private if for all S ⊆ Range($\mathcal{M}$) and for all $x, y \epsilon \mathbb{N}^{|\mathcal{X}|}$ such that $||x - y||_1 \leq 1$ will follow:
$$P(\mathcal{M}(x) \epsilon S) \leq \exp(\epsilon) P(\mathcal{M}(y) \epsilon S) + \delta$$

where the probability space is over the coin flips of the mechanism $\mathcal{M}$. If $\delta = 0$, we say that $\mathcal{M}$ is $\epsilon$-deferentially private.

*Distance between two adjacent databases*

The $\ell_1$ norm of a database $x$ is denoted by $||x||_1$ and is defined to be

$$||x||_1 = \sum_{i=1}^{|x|} |x|_i$$

The $\ell_1$ distance between two databases $x$ and $y$ is $||x - y||_1$ where the later term denotes the measure of how many records differ between $x$ and $y$.

*Explanation of Different DP*

The values of δ that are less than the inverse of any polynomial in the size of the database are preferable. The values of δ which are O(1/ ||x||₁) are dangerous as it allows to show all private information of individuals from a very small portion of the database like "just a few" case that we are already discussed in previous subsection. If δ is negligible, there are differences between ($\epsilon$,0) and ($\epsilon, \delta$)differential privacy. ($\epsilon$,0)-differential privacy says that, for every run of the mechanism $\mathcal{M}(x)$, the output observed is (almost) equally likely to be observed on every neighboring database, simultaneously. Whereas, ($\epsilon$,δ)-differential privacy provide us a mechanism where, for every pair of neighboring databases x, y, it is extremely unlikely that, 'ex post facto' the observed value $\mathcal{M}(x)$ will be much more or much less likely to be generated when the database is x than when the database is y.

*Intuition for Post-processing*

For an output $\mathcal{Z} \sim \mathcal{M}(x)$, a database y is may be possible to find such that $\mathcal{Z}$ is much more likely to be produced on y than it is produced when in database x. Then, privacy loss can be calculated which is given below:

$$\mathcal{L}^{\mathcal{Z}}_{\mathcal{M}(x)||\mathcal{M}(y)} = \frac{P(\mathcal{M}(x)=\mathcal{Z})}{P(\mathcal{M}(y)=\mathcal{Z})}$$

This loss can be negative or positive.

*Probability Simplex*

Given a discrete set $\mathcal{B}$, the probability simplex over $\mathcal{B}$, denoted $\Delta(\mathcal{B})$ is defined to be:

$$\Delta(\mathcal{B}) = \{x \epsilon \mathbb{R}^{|\mathcal{B}|} : x_i \geq 0 \; \forall i \; and \; \sum_{i=1}^{|\mathcal{B}|} x_i = 1\}$$

*Randomized Algorithm*

A randomized algorithm $\mathcal{M}$ with domain $\mathcal{A}$ and discrete range $\mathcal{B}$ is associated with a mapping $\mathcal{M} : \mathcal{A} \rightarrow \Delta(\mathcal{B})$. On input $a \epsilon \mathcal{A}$, the algorithm $\mathcal{M}$ outputs $\mathcal{M}(a) = b$ with probability $(\mathcal{M}(a))_b$ for each $b \epsilon \mathcal{B}$. The probability space is over the coin flips of the algorithm $\mathcal{M}$.

## Theorem

Any ($\epsilon$,δ)-deferentially private mechanism $\mathcal{M}$ is (k$\epsilon$ ,δ) differentially private for groups of size k. That is, for all $||x − y||1 \leq k$ and all S $\subseteq$ Range$\mathcal{M}$]

$$P[\mathcal{M}(x) \epsilon S] \leq \exp(k\epsilon) P[\mathcal{M}(y) \epsilon S]$$

where the probability space is over the coin flips of the mechanism $\mathcal{M}$.

*Definitions and Theorems for the second step:*

Here we will discuss some useful definitions and background theorems that work for the second learning step of DP discussed in the useful background on differential privacy section in alphabetical order as instructed by the publisher (Dwork et. al., 2014).

## Theorems

*Additive Chernoff's bound*

Let $\mathcal{X}_1, \ldots, \mathcal{X}_m$ be independent random variables bounded such that $0 \leq \mathcal{X}_i \leq 1$ for all $i$. Let $S = \frac{1}{m}\sum_{i=1}^{m}\mathcal{X}_i$ denote their mean and let $\mu = \mathbb{E}[S]$ denote their expected mean. Then:

$$P[S > \mu + \varepsilon] \leq e^{-2m\varepsilon^2}$$
$$P[S < \mu - \varepsilon] \leq e^{-2m\varepsilon^2}$$

*Azuma's Inequality*

Let $f$ be a function of $m$ random variables $\mathcal{X}_1, \ldots, \mathcal{X}_m$ each $\mathcal{X}_i$ taking values from a set $A_i$ such that $\mathbb{E}[f]$ is bounded. Let $c_i$ denote the maximum effect of $\mathcal{X}_i$ on $f$, then for all $a_i, \acute{a}_i \in A_i$:

$$|\mathbb{E}[f \mid \mathcal{X}_1, \ldots, \mathcal{X}_{i-1}, \mathcal{X}_i = a_i] - \mathbb{E}[f \mid \mathcal{X}_1, \ldots, \mathcal{X}_{i-1}, \mathcal{X}_i = \acute{a}_i]| \leq c_i$$

Then:

$$P[f(\mathcal{X}_1, \ldots, \mathcal{X}_m) \geq \mathbb{E}[f] + t] \leq e^{-\left(\frac{2t^2}{\sum_{i=1}^{m} c_i^2}\right)}$$

*Exponential Mechanism*

The exponential mechanism $\mathcal{M}_E(x, u, \mathcal{R})$ selects and outputs an element $r\epsilon\mathcal{R}$ with probability proportional to $\exp(\frac{\epsilon u(x,r)}{2\Delta u})$. Where, sensitivity of the utility score u: $\mathbb{N}^{|\mathcal{X}|} \times \mathcal{R} \to \mathbb{R}$ and

$$\Delta u = \max_{r \in \mathcal{R}} \max_{x,y \in \|x-y\| \leq 1} \|u(x,r) - u(y,r)\|$$

*Claim*

The Exponential Mechanism holds ($\epsilon$,0) differential privacy.

*Gaussian Mechanism*

Just like Laplace Mechanism, it adds noise drawn from the Gaussian distribution whose variance is calibrated according to the sensitivity and privacy parameters.

$$\mathcal{M}_{Gauss}(x, f, \epsilon, \delta) = f(x) + \mathcal{N}^d\left(\mu = 0, \sigma^2 = \frac{2\ln\left(\frac{1.25}{\delta}\right).(\Delta_2 f)^2}{\epsilon^2}\right)$$

*Claim*

$\mathcal{M}_{Gauss}$ holds ($\epsilon$,$\delta$) differential privacy

*Multiplicative Chernoff's bound*

Here the background assumption is same as the additive Chernoff's bound. The bound is following:

$$P[S > \mu(1 + \varepsilon)] \leq e^{-2m\varepsilon^2/3}$$
$$P[S < \mu(1 - \varepsilon)] \leq e^{-2m\varepsilon^2/2}$$

*Noisy Max Report*

If we consider a simple algorithm where it determines which m counting queries has highest value, and then it adds generated Laplace Noise from Lap($1/\epsilon$) to each count and return the index of the largest noisy count. We can ignore the possibilities of tie. This is called Noisy Max Report.

> *Claim*
>
> The Report Noisy Max algorithm is ($\epsilon$,0)-differentially private.
>
> *Report One-Sided Noisy Arg-Max*
>
> When run with parameter $\epsilon/2\delta u$ yields the same distribution on outputs as the exponential mechanism.

## Randomized Response

It's a simple survey process where binary responses are considered i.e. the answers should be in 'yes' and 'no' only. The steps are following:
- Toss a coin
- If the coin shows 'tails', then give the 'true' answer
- If the coin shows 'heads', then again toss it and give 'yes' if the result is 'heads' or 'no' if the result is 'tails'.

For ($\epsilon$,$\delta$) differentially private mechanism, to find the value of $\epsilon$, we obtain:

$$RR = \frac{P[Response=\text{YES}|Truth=\text{"YES"}]}{P[Response=\text{"YES"}|Truth=\text{"NO"}]}$$

where, $\epsilon = \ln(RR)$.

> *Claim*
>
> Randomized Response Technique holds (,0) differential privacy where $\epsilon = \ln(3)$. For more analysis, remarks and corollary please refer to (Dwork et. al., 2014).

## Stirling's Approximation

$n!$ Can be approximated by $\sqrt{2\pi n}(\frac{n}{e})^n$.

## The Laplace Mechanism

Given any function f: $\mathbb{N}^{|\mathcal{X}|} \to \mathbb{R}^k$, the Laplace mechanism is defined as:

$$\mathcal{M}_L(x, f(.), \epsilon) = f(x) + (Y_1, Y_2, \ldots\ldots, Y_k)$$

Where $Y_i$ ; $i = 1{:}k$, are i.i.d random variables drawn from Lap($\Delta f/\epsilon$).

## Definitions and Theorems for the third step:

Here we will discuss some useful definitions and background theorems that work for the third learning step of DP discussed in the useful background on differential privacy section in alphabetical order as instructed by the publisher (Dwork et. al., 2014).

## Theorems

### Accuracy

We will say that an algorithm which outputs a stream of answers $a_1, \ldots, \epsilon(\top, \bot)^*$ in response to a stream of k queries $f_1, \ldots, f_k$ is ($\alpha$,$\beta$)-accurate with respect to a threshold $\bot$ if except with probability at most β, the algorithm does not halt before $f_k$, and
$\forall a_i = \top$:

$$f_i(D) \geq T - \alpha.$$

and, $\forall a_i = \bot$:

$$f_i(D) \leq T + \alpha$$

## Advanced Composition

If $\mathcal{A}_1, \ldots, \mathcal{A}_k$ are randomised algorithm satisfying $(,\delta)$ differential privacy, then their composition, defined as $\mathcal{A}_1(D), \ldots, \mathcal{A}_1(D), for D \epsilon \mathcal{D}$ satisfies $(\epsilon', k\delta + \delta')$ differential privacy where $\epsilon' = \epsilon\sqrt{2klog(1/\delta')} + k\epsilon(\exp(\epsilon) - 1)$. Moreover $\mathcal{A}_i$ can be chosen adaptively depending on the outputs of $\mathcal{A}_1, \ldots, \mathcal{A}_k$.

Let, two distribution are given $\mu$ and $\mu'$, then we can say that they are $(\epsilon, \delta)$- deferentially close, denoted by $\mu \cong_{(\epsilon,\delta)} \mu'$, if for all measurable $\mathcal{A}$, we have

$$\exp(-\epsilon)(\mu'(\mathcal{A} - \delta)) \leq \mathcal{A} \leq \exp(\epsilon)(\mu'(\mathcal{A} + \delta))$$

## Composition

Let $\mathcal{M}_i: \mathbb{N}|X| \to \mathcal{R}_i$ be an $(\epsilon_i, \delta_i)$ differentially private algorithm for i $\epsilon$ [k]. Then if $\mathcal{M}_{[k]} : \mathbb{N}^{|\mathcal{X}|} \to \prod_{i=1}^{k} \mathcal{R}_i$ is defined to be $\mathcal{M}_{[k]}(x) = (\mathcal{M}_1(x), \mathcal{M}_2(x), \ldots, \mathcal{M}_k(x))$, then $\mathcal{M}_{[k]}(x)$ is $(\sum_{i=1}^{k} \epsilon_i, \sum_{i=1}^{k} \delta_i)$ −differentially private.

### Corollary 1

Let $\mathcal{M}_i : \mathbb{N}|X| \to \mathcal{R}i$ be an $(\epsilon_i, 0)$ differentially private algorithm for i $\epsilon$ [k]. Then if $\mathcal{M}_{[k]} : \mathbb{N}^{|\mathcal{X}|} \to \prod_{i=1}^{k} \mathcal{R}_i$ is defined to be $\mathcal{M}_{[k]}(x) = (\mathcal{M}_1(x), \mathcal{M}_2(x), \ldots, \mathcal{M}_k(x))$, then $\mathcal{M}_{[k]}(x)$ is $(\sum_{i=1}^{k} \epsilon_i, 0)$ − differentially private.

## Properties of Differential Privacy

The notion of $(\epsilon, \delta)$ differential private satisfies the following properties:

- **Monotonicity**: Let. $\mu \cong_{(\epsilon,\delta)} \mu'$ Then for $\epsilon' > \epsilon$ and $\delta' > \delta, \mu \cong_{(\epsilon',\delta')} \mu'$.
- **Triangle Inequality** : Let, $\mu_1 \cong_{(\epsilon_1,\delta_1)} \mu_2$ and $\mu_2 \cong_{(\epsilon_2,\delta_2)} \mu_3$, then $\mu_1 \cong_{(\epsilon_1+\epsilon_2,\delta_1+\delta_2)} \mu_3$
- **Quasi-Convexity**: Let, $\mu_1 \cong_{(\epsilon,\delta)} \mu_1'$ and, $\mu_2 \cong_{(\epsilon,\delta)} \mu_2'$ then for any a $\epsilon$ [0,1], it holds that $(1-a)\mu_1 + a\mu_2 \cong_{(\epsilon,\delta)} (1-a)\mu_1' + a\mu_2'$.

### Lemma 1

Let $q < \frac{1}{2}$ and let $\mu_0, \mu_1$ be distributions such that $\mu_1 \cong_{(\epsilon,\delta)} \mu_0$. For $\mu = (1-q)\mu_0 + q\mu_1$, it holds that $\mu \cong_{(\epsilon',q\delta)} \mu_0$, where $\epsilon' = \log(q(e^\epsilon - 1) + 1) \leq q(e^\epsilon - 1)$.

## Max Divergence

The Max Divergence between two random variables Y and Z taking values from the same domain is defined to be:

$$D_\alpha(Y||Z) = \max_{S \subseteq \sup(Y)} [ln \frac{P[Y \in S]}{P\{Z \in S\}}]$$

The δ-Approximate Max Divergence between Y and Z is defined to be:

$$D_\alpha^\delta(Y||Z) = \max_{S \subseteq \sup(Y): P(Y \in S) \geq \delta} [\ln \frac{P[Y \in S] - \delta}{P[Z \in S]}]$$

### Remark 1

For a mechanism $\mathcal{M}$, it is to be noted:

- It will be differentially private necessary and sufficiently for two neighbouring database $x$ and $y$, $D_\alpha(\mathcal{M}(x)||\mathcal{M}(y)) \leq \epsilon$ and $D_\alpha(\mathcal{M}(y)||\mathcal{M}(x)) \leq \epsilon$; and
- It will be $(\epsilon, \delta)$ private necessary and sufficiently for two neighbouring data set $x$ and $y$, $D_\alpha^\delta(\mathcal{M}(x)||\mathcal{M}(y)) \leq \epsilon$ and $D_\alpha^\delta(\mathcal{M}(y)||\mathcal{M}(x)) \leq \epsilon$.

## $l_1$-Sensitivity

The $l_1$ sensitivity of a function f : $\mathbb{N}^{|\mathcal{X}|} \to \mathbb{R}^k$, then: $\Delta f = \max\limits_{x,y \in \mathbb{N}^{|\mathcal{X}|}, ||x-y||_1 = 1} ||f(x) - f(y)||_1$

## $l_2$-Sensitivity

The $l_2$ sensitivity of a function f : $\mathbb{N}^{|\mathcal{X}|}$ then: $\Delta_2 = \max\limits_{x,y \in \mathbb{N}^{|\mathcal{X}|}} ||f(x) - f(y)||_2$.

## Utility

How much information about the real answer can be obtained from the reported one by a specific model[18], is called the utility. How the utility works can be defined by the figure 2.